\documentclass[11pt]{article}
\usepackage{graphicx}
\usepackage{cancel}[]
\usepackage{amssymb,amsmath}
\usepackage{ifpdf}
\usepackage{fullpage}
\usepackage{color}
\usepackage{ulem}
\ifpdf
    \DeclareGraphicsExtensions{.pdf}
\else
    \DeclareGraphicsExtensions{.eps}
\fi

\begin{document}

\begin{center}
\vspace{0.3in} {\large {\bf{The Effect of the Tides on the LIGO
Interferometers}}}
\\

\vspace{0.1 in} \rm{ A. C. Melissinos, for the LIGO
Scientific Collaboration}
\\

\vspace{0.1 in}

\it{ Department of Physics and Astronomy, University of Rochester,
\\
Rochester NY 14627, USA} \rm

\vspace{0.1 in} \rm{To appear in the Proceedings of the 12th Marcel
Grossman Meeting, Paris, July 2009}\\
\rm{ LIGO Document P0900257-v3}
 \end{center}

\vspace{0.2 in}

The LIGO interferometers \cite{LIGO} can record the dark port signal
at a sampling rate as high as 262.144 kHz. Of interest is the region
around the 1st free spectral range (fsr) of the 4 km
interferometers, $f_{fsr} = c/2L = 37.52$ kHz. The dominant optical
field in the interferometer, $E_0$, is at the carrier frequency
$f_0$ and has narrow width $\Delta f_0 \sim 1-2$ Hz. In addition in
the interferometer circulate sideband fields $E{_\pm}(f)$ centered
at $f_{\pm} = f_0 \pm f_{fsr}$. The sideband fields have width
typical of the arm cavity resonance, $\Delta f_{\pm} \sim 200$ Hz,
and peak amplitude density $E_{\pm}(f) \sim 10^{-7} E_0
/\sqrt{\rm{Hz}}$.
%\cite{Maryland,parametric}
When the interferometer is locked, the field $E_0$ is on a dark
fringe, and  at the antisymmetric port (AS), the phase shift at the
carrier frequency, $\Delta \phi_0 = 0$. However this does not hold
for the sideband fields $E_{\pm}(f)$ because the lengths $L_x$ and
$L_y$ of the two arms are not equal. Instead  the phase shift at
frequency $f_{\pm} = f_0 \pm f_{fsr}$ is finite and given by $
\Delta \phi_{\pm} = \pm 2\pi (L_x - L_y)/L$, where $L_x - L_y
 \sim 2$ cm.
The fields $E_{\pm}(f)$ are mixed with the radio frequency
sidebands, and the photocurrent is demodulated in the usual way.
This gives rise to a signal centered at $f_{fsr}$ and of amplitude
density $h(f)$ proportional to the phase shift $\Delta \phi_{\pm}$
and the field density $E_{\pm}(f)$, $ h(f) = C |\Delta
\phi_{\pm}||E_{\pm}(f)|$. The uncalibrated frequency spectrum in
counts/$\sqrt{\rm{Hz}}$ , in the region of the fsr frequency is
shown in Fig.1. %\cite{Stefanos,LSU}
 The enhancement follows the spectral shape of $E_{\pm}(f)$.\\

The data are available as a time series sampled at 1024 Hz and
restricted to the frequency range $37,504 \pm 1024$ Hz. The data are
from the S5 science run and were written in 64 s long frames. For
each frame we obtain the frequency spectrum at a resolution BW =
0.125 Hz %\cite{fsr,Stefanos}
 and integrate the power spectral
density (PSD) in the region $f_{fsr} \pm 200$ Hz. A filter
proportional to the interferometer transfer function at the fsr is
applied. %\cite{Chad}.
This gives a time series of the power at the fsr, sampled every 64
s. The amplitude of this signal is proportional to the sum of the
``biasing" phase shift $\Delta \phi_{\pm}$ given above, and of any
time-dependent phase shift $\Delta \phi_t$ at the sideband
frequencies $f_{\pm}$. Thus the observed (integrated) power is

\begin{equation}
P \ = \int{(\rm{PSD})} df = \big\{(\Delta \phi_{\pm})^2 + 2\Delta
\phi_{\pm}\Delta \phi_t + (\Delta \phi_t)^2\big\}|C|^2
\int|E_{\pm}(f)|^2 df
\end{equation}

\noindent Since $\Delta \phi_t$ is less than few percent of $\Delta
\phi_{\pm}$, the last term is negligible and the time-dependent
modulation of the power is

\begin{equation}
\Delta P/P = 2 |\Delta \phi_t|/|\Delta \phi_{\pm}|
\end{equation}

\noindent Measurement of $\Delta P$ combined with a knowledge of
$\Delta \phi_{\pm}$, (i.e. of $\Delta L$), suffices to obtain
$\Delta \phi_t$.\\

 A plot of the power for a 16 month period (April
2006 to August 2007) for the Hanford H1 interferometer is shown in
Fig.2. The modulation at the daily and twice daily frequencies is
clearly seen in the inset. The 16-month time series was spectrally
analyzed and contains the tidal frequencies as shown in Figs 3(a,b).
To within the measurement resolution of $\Delta f_{res} = 6\times
10^{-9}$ Hz there is exact agreement between the measured and
expected values. The modulation of the power is  shown as a function
of time for a five day period in December 2006 in Fig.4a, and for
April 2007 in Fig.4b. The blue curves give the relative amplitude of
the horizontal component of the tidal forces at the Hanford site for
the same time period.

%\end{document}

\begin{figure} [h!]
\centering
\includegraphics[width=115mm,height=95mm]{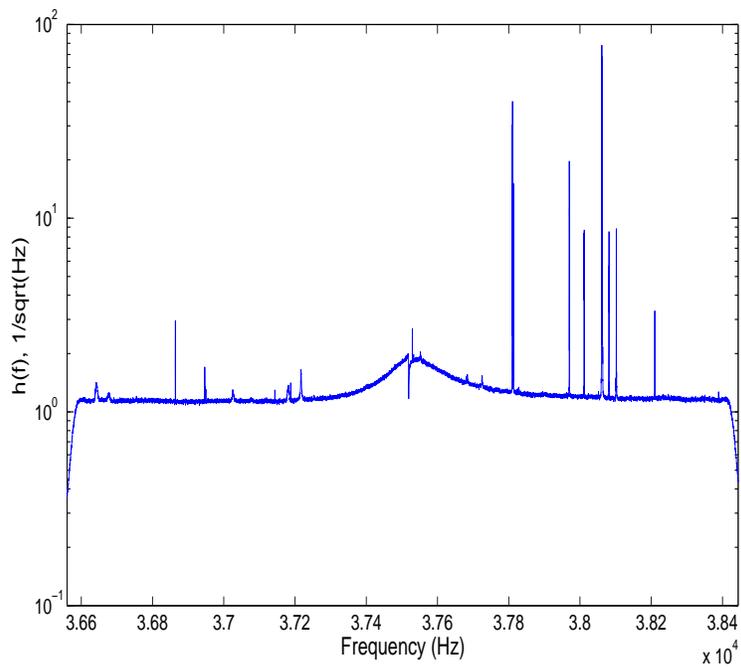}
\caption{ Uncalibrated amplitude spectral density for the H1
interferometer in the region of its free spectral range.}
\end{figure}

%\newpage

\begin{figure}
%\centering
\includegraphics[width=150mm,height=100mm]{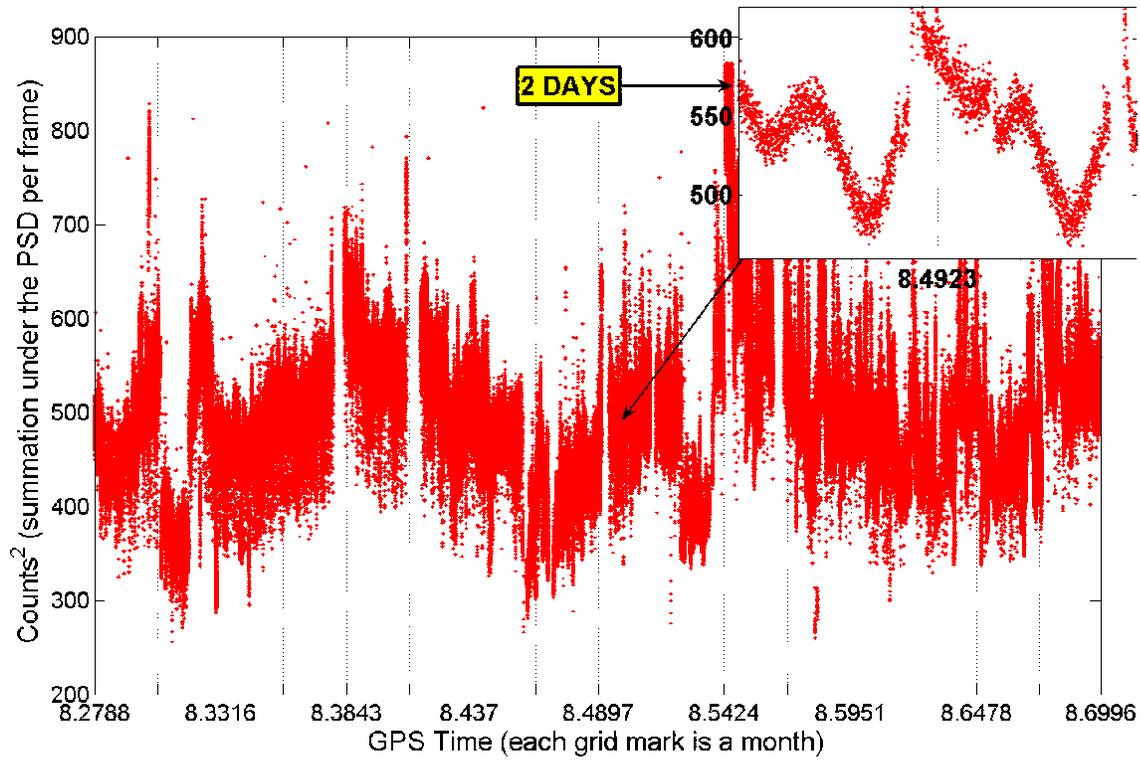}
\caption{Integrated power in the free spectral range (fsr) region as
a function of time, April 2006 to July 2007. The data are for the H1
interferometer and are sampled every 64 s. Note the daily and
twice-daily modulation that can be seen in the inset.}
\end{figure}

%\begin{figure} [h!]
%\centering
%\includegraphics[width=130mm,height=90mm]{Power(spectrum)_200_tf_Apr06_Jul07}
%\caption{Top plot shows an uncalibrated spectrum for H1:LSC-AS\_Q}
%\end{figure

\newpage

\begin{figure} [h!]
\centering %\subfigure[]{
\includegraphics[width=115mm,height=75mm]{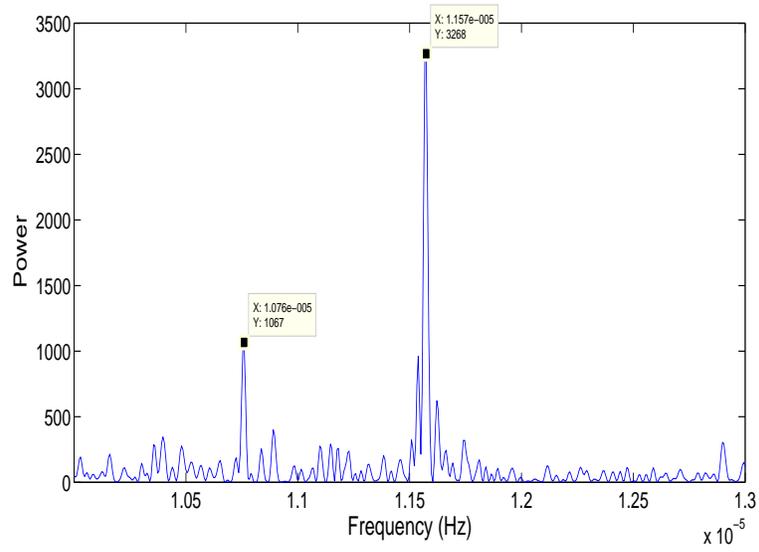}
%\subfigure[]{
\caption{Frequency spectrum of the integrated fsr power in the
diurnal region. Note the fine structure.}
\end{figure}

\begin{figure} [h!]
\centering
\includegraphics[width=115mm,height=75mm]{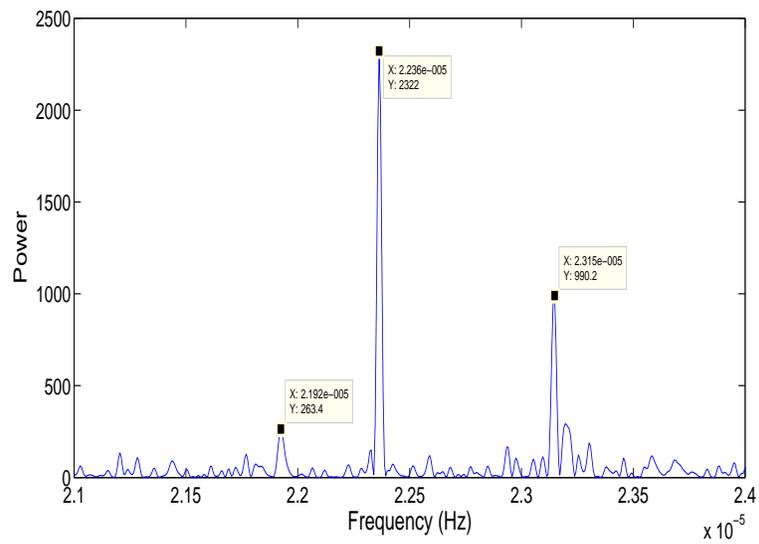}
\caption{Frequency spectrum of the integrated fsr power in the twice
daily region. Note the fine structure.}
\end{figure}

\begin{figure} [h!]
\centering %\subfigure[]{
\includegraphics[width=115mm,height=75mm]{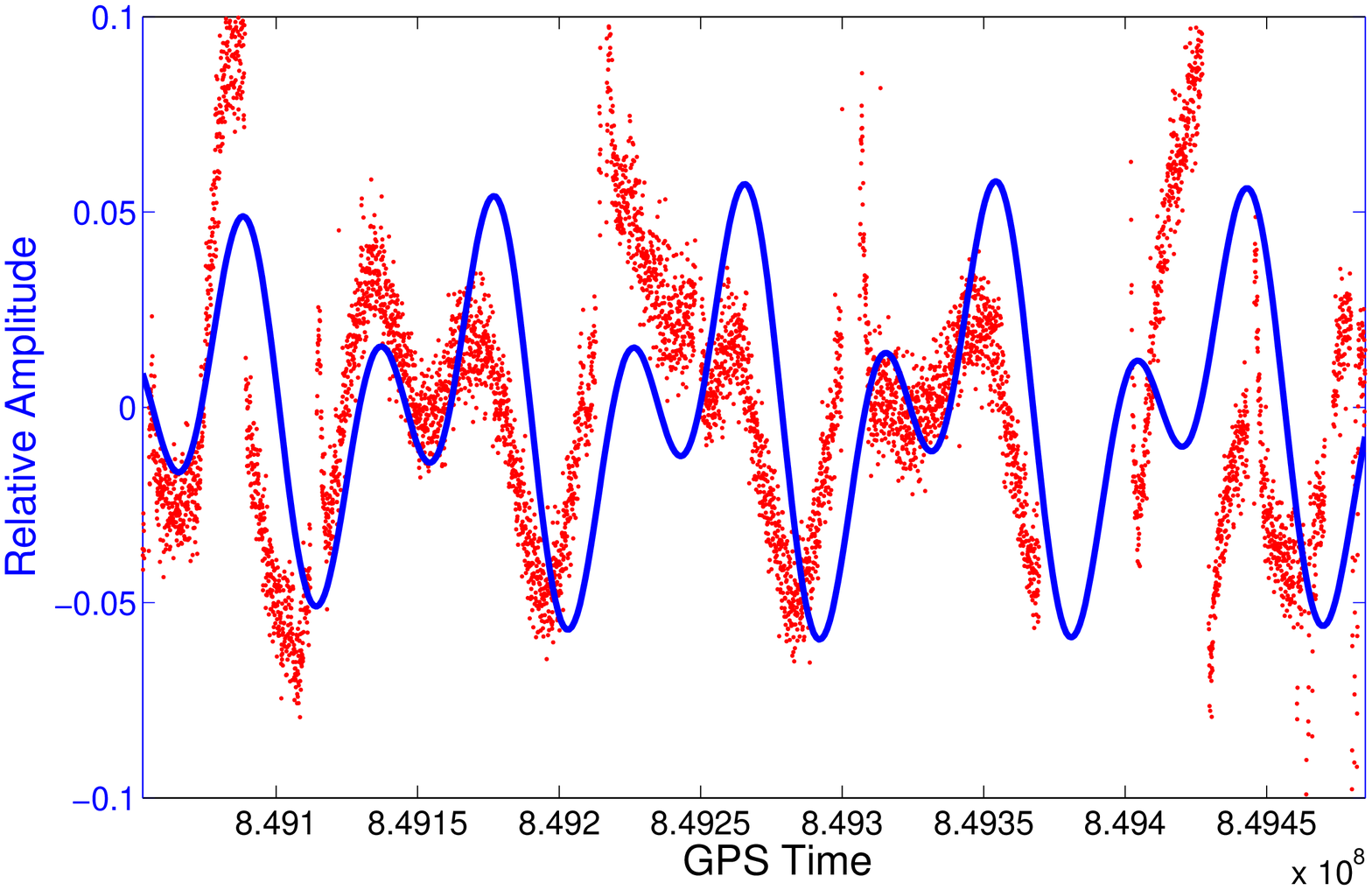}
%\subfigure[]{
\caption{Modulation of the integrated fsr power for five days in
December 2006. The blue curves give the relative amplitude of the
horizontal component of the tidal forces at the Hanford site for the
same time period.}
\end{figure}

\begin{figure} [h!]
\centering
\includegraphics[width=115mm,height=75mm]{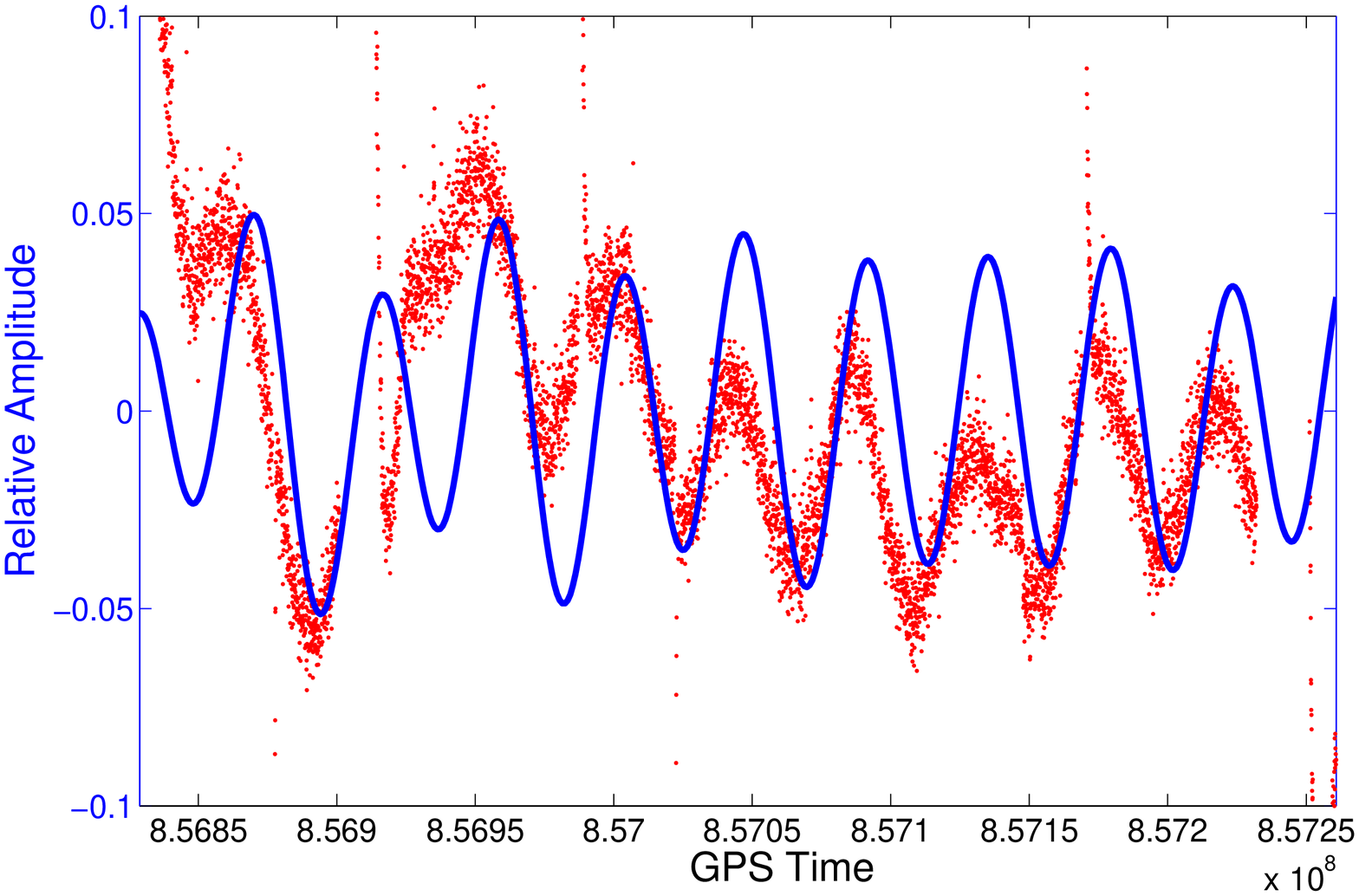}
\caption{Modulation of the integrated fsr power for five days in
March 2007. The blue curves give the relative amplitude of the
horizontal component of the tidal forces at the Hanford site for the
same time period.}
\end{figure}


\begin{thebibliography}{99}

\bibitem{LIGO} B.Abbott et al. Nucl. Instrum. Methods, A{\bf{517}}
154 (2004).


\end{thebibliography}
\end{document}